\documentclass[preprint]{aastex}

\def\kms{~km~s$^{-1}$}

\def\hal{H$\alpha$}
\def\be{\begin{equation}}
\def\ee{\end{equation}}

\def\m{~$\mu$m}

\def\HII{\ion{H}{2}}

\def\Htwo{H$_2$}

\def\ISO{{\it ISO}}
\def\IRAS{{\it IRAS}}

\def\IRAScolor{${f_\nu (60 \mu {\rm m})} \over {f_\nu (100 \mu {\rm m})}$}

\def\colore{${f_\nu (6.75 \mu {\rm m})} \over {f_\nu (15 \mu {\rm m})}$}
\def\COa{$^{12}$CO(1$\rightarrow$0)}

\begin {document}

\title{Warm and Cold Molecular Gas in Galaxies}
 
\author{Daniel A. Dale,\altaffilmark{1} Kartik Sheth,\altaffilmark{2} George~Helou,\altaffilmark{2,3} Michael~W.~Regan\altaffilmark{4} and Susanne~H\"uttemeister\altaffilmark{5}}
\altaffiltext{1}{\scriptsize Department of Physics and Astronomy, University of Wyoming, Laramie, WY 82071; ddale@uwyo.edu}
\altaffiltext{2}{\scriptsize Spitzer Science Center, California Institute of Technology, M.S. 220-6, Pasadena, CA 91125}
\altaffiltext{3}{\scriptsize Infrared Processing and Analysis Center, California Institute of Technology, M.S. 1000-22, Pasadena, CA 91125}
\altaffiltext{4}{\scriptsize Space Telescope Science Institute, 3700 San Martin Drive, Baltimore, MD 21218}
\altaffiltext{5}{\scriptsize Astronomisches Institut der Ruhr-Universit\"at Bochum, Universit\"atsstr. 150, 44780 Bochum, Germany}

\begin{abstract}
New and archival interferometric \COa\ datasets from six nearby galaxies are combined with \Htwo~2.122\m\ and \hal\ maps to explore in detail the interstellar medium in different star-forming galaxies.  We investigate the relation between warm (\Htwo\ at $T\sim2000$~K) and cold (CO at $T\sim50$~K) molecular gas from 100~pc to 2~kpc scales.  On these scales, 
the ratio of warm-to-cold molecular hydrogen correlates with the \IRAScolor\ ratio, a ratio that tracks the star formation activity level.  This result also holds for the global properties of galaxies from a much larger sample drawn from the literature.  The trend persists for over three orders of magnitude in the mass ratio, regardless of source nuclear activity.
\end{abstract}

\keywords{galaxies: ISM --- infrared: galaxies --- radio lines: galaxies}

\section{INTRODUCTION}
Carbon monoxide is an important probe of the interstellar medium.  It is used to map large- and small-scale  kinematic signatures within galaxies, it traces the sites of possible future star formation, and it is a favored proxy for estimating the amount of \Htwo.  \Htwo\ is the predominant interstellar molecular species but it is more difficult to detect than CO.  Researchers frequently use ratios of various radio transitions of CO to gauge the physical conditions of the interstellar medium.  Such transitions reflect cool gas at temperatures of order $5-20$~K for disks and $50-75$~K for nuclear regions (e.g. Young \& Scoville 1991; Harris et al. 1991; Hafok \& Stutzki 2003).  The mid-infrared spectroscopic capabilities of the {\it Infrared Space Observatory} enabled observations of warm \Htwo\ gas ($T\sim150-700$~K) via emission that arises from pure rotation of the molecule (e.g. Timmermann et al. 1996; Valentijn \& van der Werf 1999; Fuente et al. 1999; Rigopoulou et al. 2002).  Rigopoulou et al. found that the mass fraction of this warm molecular gas, with respect to the total molecular gas as traced by \COa, is 1-10\% for starbursts and 2-35\% for Seyferts (see also Armus et al. 2004; Rodr\'{\i}guez-Fern\'andez et al. 2004).  The fraction of molecular material at high temperatures ($T\sim2000$~K) indicates the amount of molecular gas either currently participating in star formation or experiencing its immediate after effects such as shocks (e.g. Scoville et al. 1982; Shull \& Beckwith 1982).  A useful probe of molecular gas at high excitation is the ro-vibrational \Htwo~2.122\m\ line.  \Htwo~2.122\m\ emission derives from collisional excitation by shocks or radiative excitation in ultraviolet radiation fields (e.g. Goldader et al. 1997), or possibly from reprocessed X-ray photons (Maloney, Hollenbach, \& Tielens 1996).  In this work we combine \COa\ and \Htwo~2.122\m\ data from six nearby galaxies to estimate the fraction of molecular material at relatively high excitation.

The sample is drawn from the large collection of galaxies studied in the $ISO$ Key Project on the Interstellar Medium in Normal Galaxies (Helou et al. 1996; Dale et al. 2000).  The data obtained for the Key Project sample include mid-infrared imaging and spectrophotometry, far-infrared spectroscopy, and \hal\ imaging.  As a follow-up to the Key Project we have recently obtained 5\farcs7$\times$10\arcsec\ \Htwo~2.122\m\ spectral data cubes and arcminutes-scale \COa\ datasets for six galaxies.  The properties of these particular galaxies are fully described in Dale et al. (2004) along with the \hal\ and near-infrared molecular hydrogen data obtained for that work.  The targets for this study span a range of environments from \HII\ regions to normal galaxy nuclei and perhaps one heavily obscured AGN.  Our maps of \Htwo~2.122\m\ emission sample sub-kpc to kpc sizescales, so coupling those data with high resolution interferometric \COa\ maps enables an exploration of hot and cold molecular gas on a local scale.  These small sizescale data are then compared to the {\it disk-averaged} properties of galaxies drawn from the literature.  

\section{\COa\ OBSERVATIONS AND DATA REDUCTION}

The \COa\ observations for NGC~1569, NGC~6946, and UGC~2855 are described in Taylor et al. (1999), Helfer et al. (2003), and H\"uttemeister et al. (1999), respectively.  The new data obtained for this work are the \COa\ maps for NGC~2388, NGC~4418, NGC~7771 described below.
 
\subsection{Owens Valley Radio Observatory CO Observations and Data Calibration}
\COa\ observations were obtained over the period 2003 February through 2003 October.  The array of six 10.4~m telescopes was used in its ``Compact'' configuration for observing one track each for NGC~2388 and NGC~7771; NGC~4418 was observed in the ``Low'' and ``Equatorial'' arrangements for one track apiece.  The average single-sideband temperature ranged from 304 to 378~K for the set of four tracks.  Spectral resolution was provided by a digital correlator configured with four bands, each with $32\times4$~MHz channels, yielding a total velocity coverage $\delta v \simeq1200$\kms\ at the redshifted \COa\ line.  

Nearby radio-loud QSOs were used for gain and passband calibration.  The data were calibrated using the standard Owens Valley array MMA program (Scoville et al. 1993) and mapped using the BIMA MIRIAD package (Sault, Teuben, \& Wright 1995).  The beam sizes for the different observations are listed in Table~\ref{tab:obs} along with the rms noise level in the source-free channels of the cleaned maps.  The planets Uranus and Neptune were used as flux calibrators; the absolute \COa\ fluxes are uncertain at the 20\% level.

\section{RESULTS}

\subsection{Morphology}
\begin{figure}[!ht]
\epsscale{1.0}
\plottwo{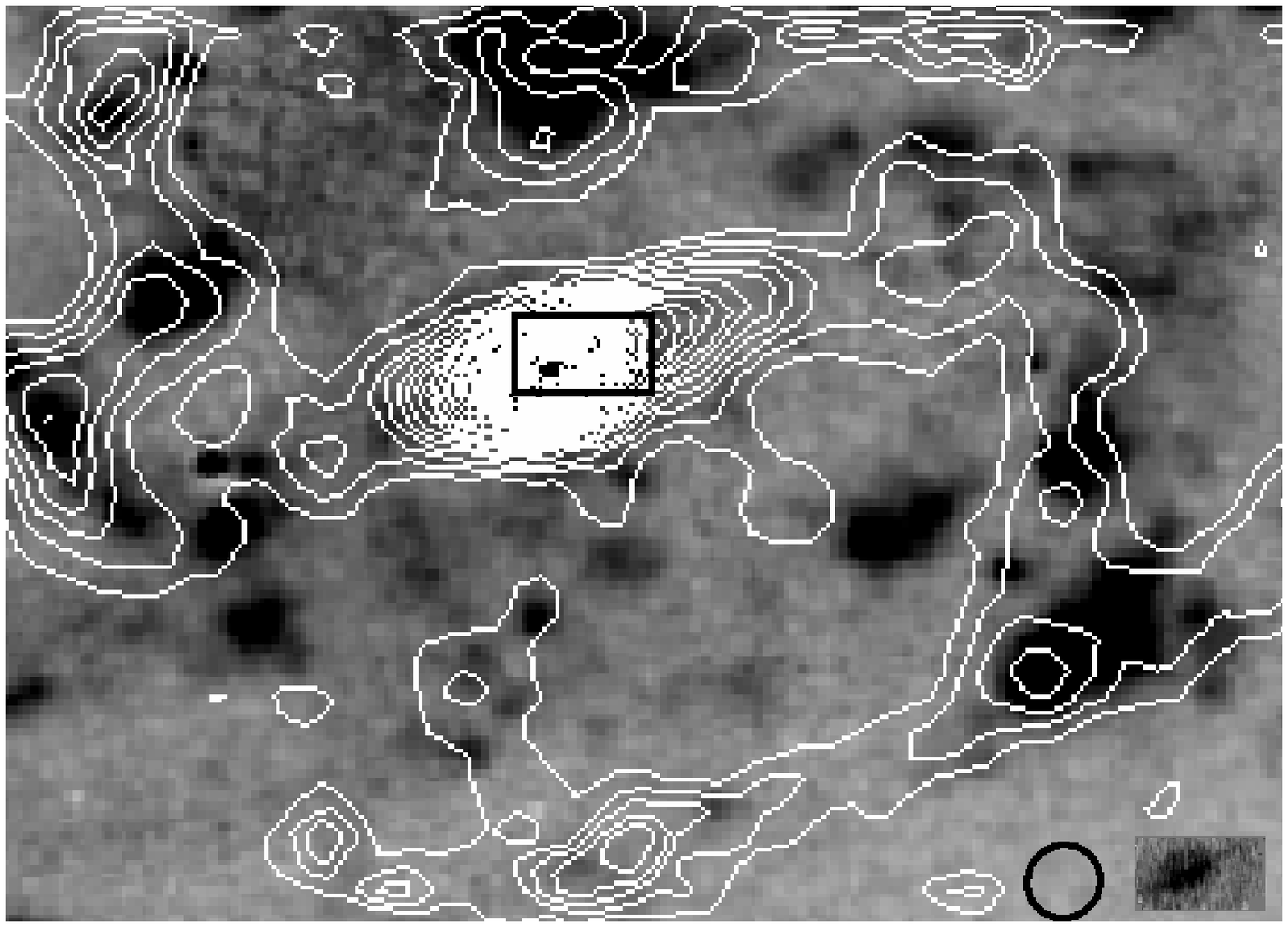}{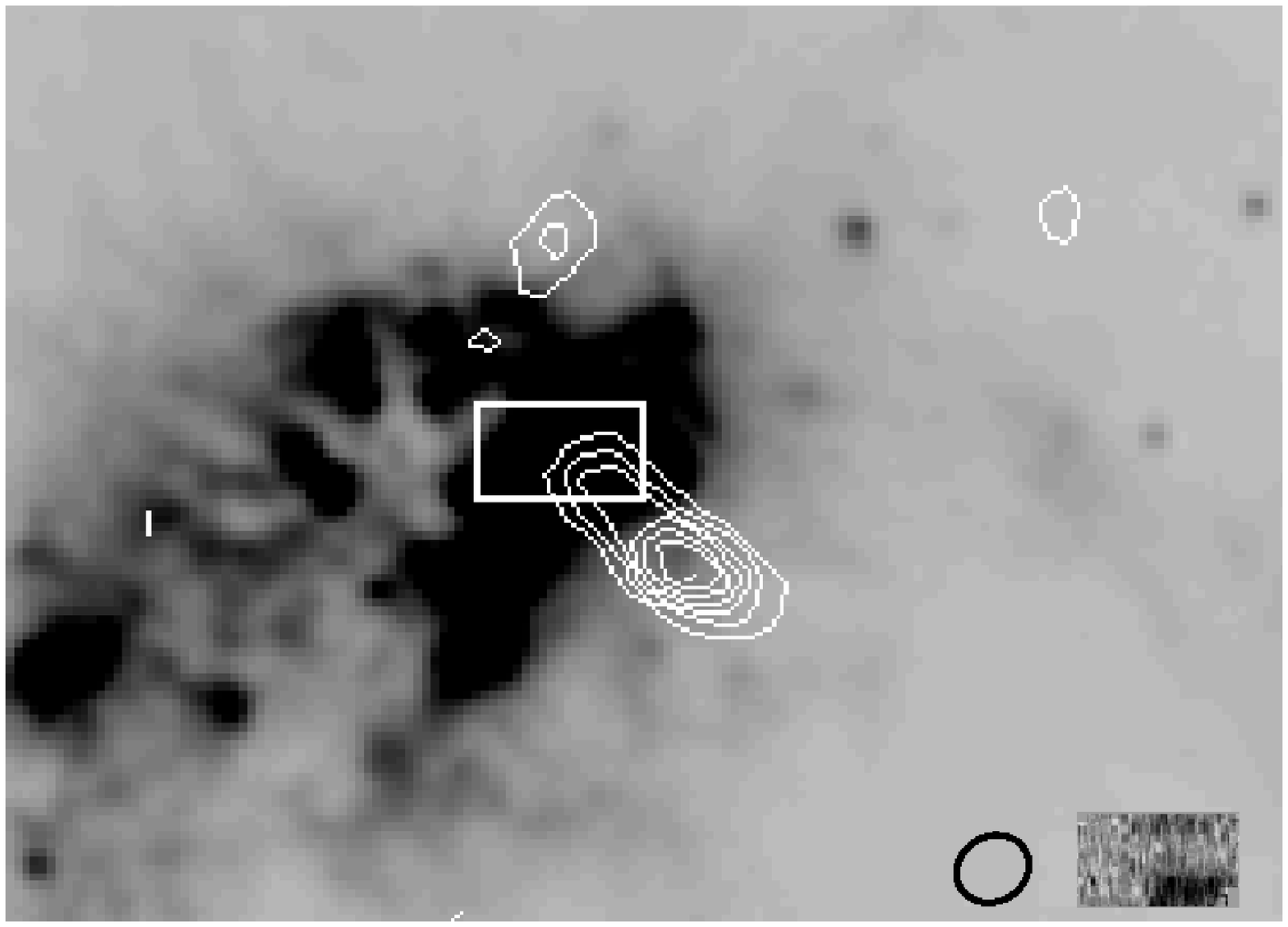}
\epsscale{1.0}
\caption[] {\ Integrated \COa\ emission contours overlaid on continuum-subtracted \hal\ images for UGC~2855 and NGC~1569.  The contours for UGC~2855 start at 3~Jy~beam$^{-1}$\kms\ and are spaced by the same value; they start at 0.2~Jy~beam$^{-1}$\kms\ and are spaced by the same value for NGC~1569.  The ellipses represent the beam sizes for the CO observations.  Rectangular outlines of the regions mapped in \Htwo~2.122\m\ are included for reference; grayscale images of the \Htwo\ maps are inset in the lower righthand corner of each image.  The \COa\ and \hal\ maps for NGC~1569 are from Taylor et al. (1999) and Martin, Kobulnicky, \& Heckman et al. (2002), respectively.    The \COa\ map for UGC~2855 is from H\"uttemeister, Aalto, \& Wall (1999).}
\label{fig:CO_ha1}
\end{figure}

\begin{figure}[!ht]
\epsscale{1.0}
\plottwo{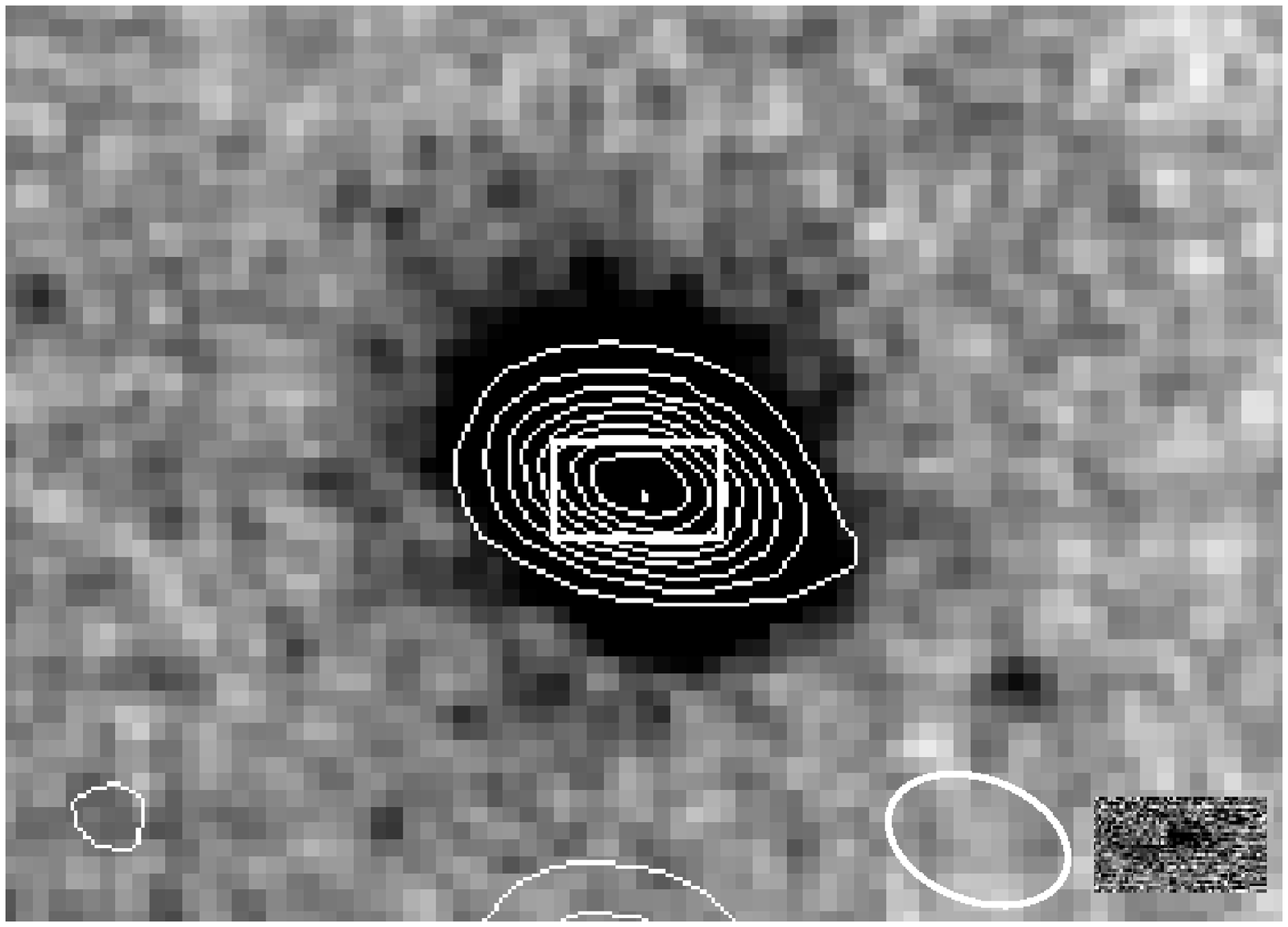}{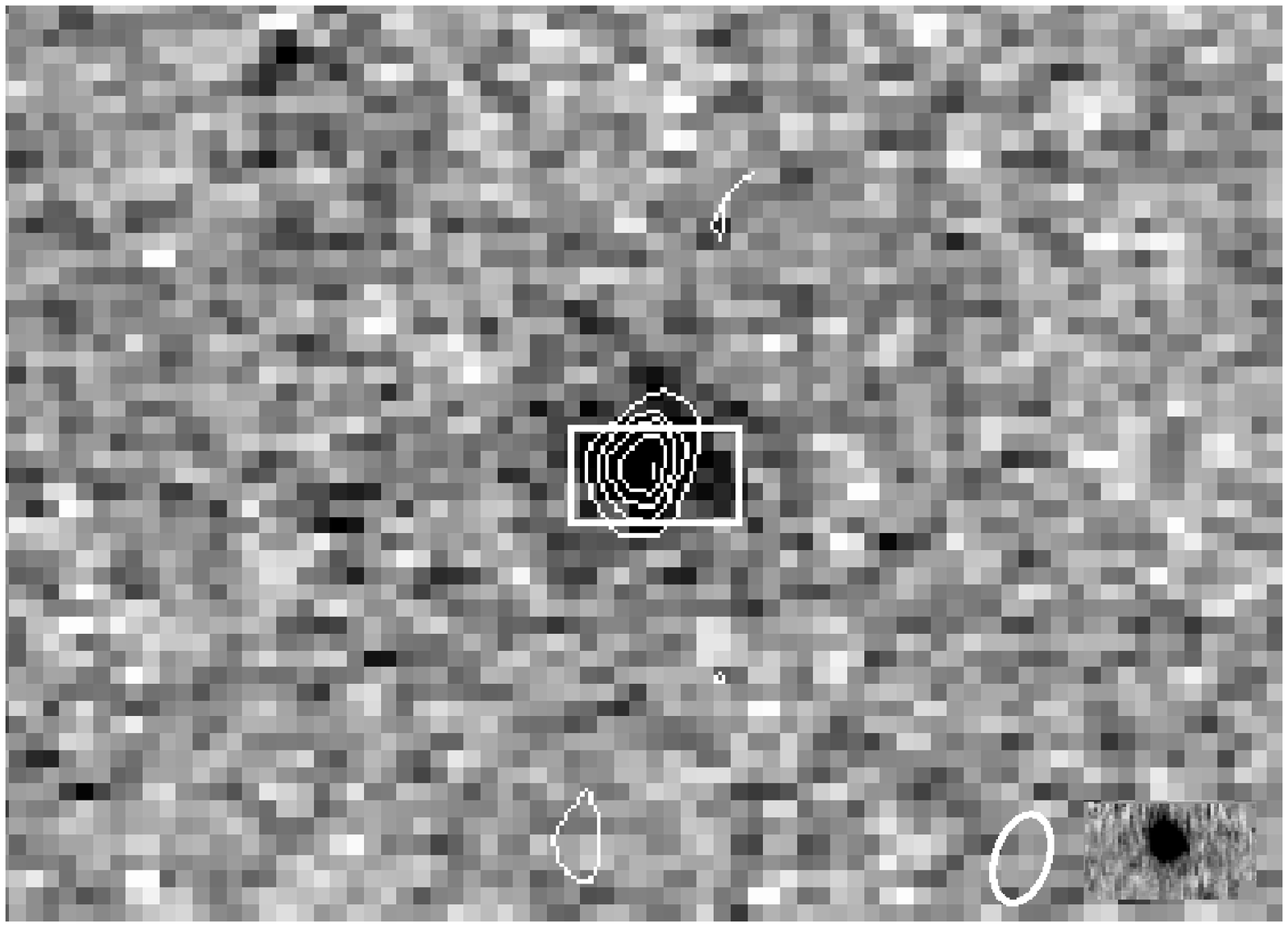}
\epsscale{1.0}
\caption[] {\ Similar to Figure~\ref{fig:CO_ha1} for NGC~2388 and NGC~4418.  The contours start at 10~Jy~beam$^{-1}$\kms\ and are spaced by the same value.} 
\label{fig:CO_ha2}
\end{figure}

\begin{figure}[!ht]
\epsscale{1.0}
\plottwo{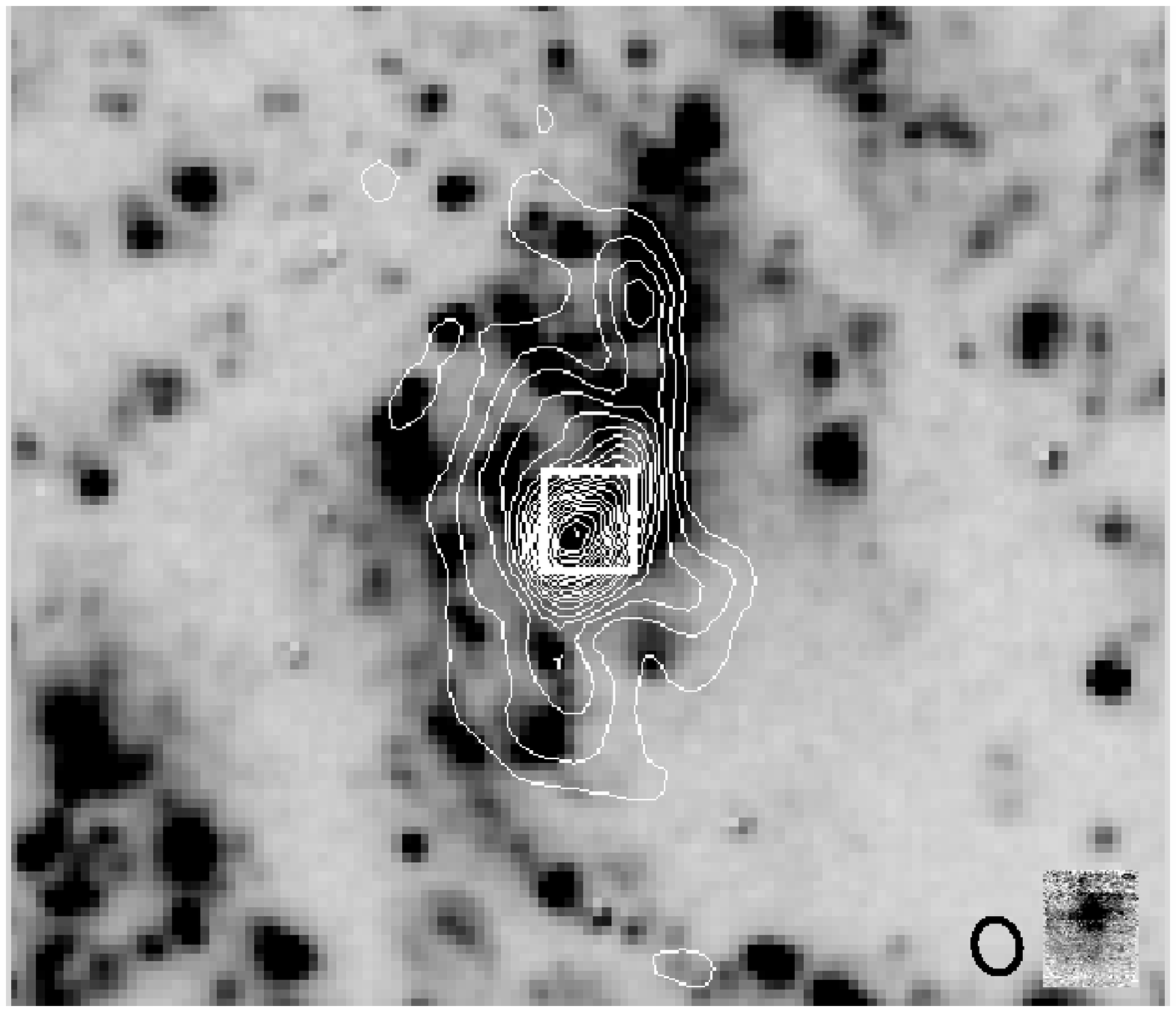}{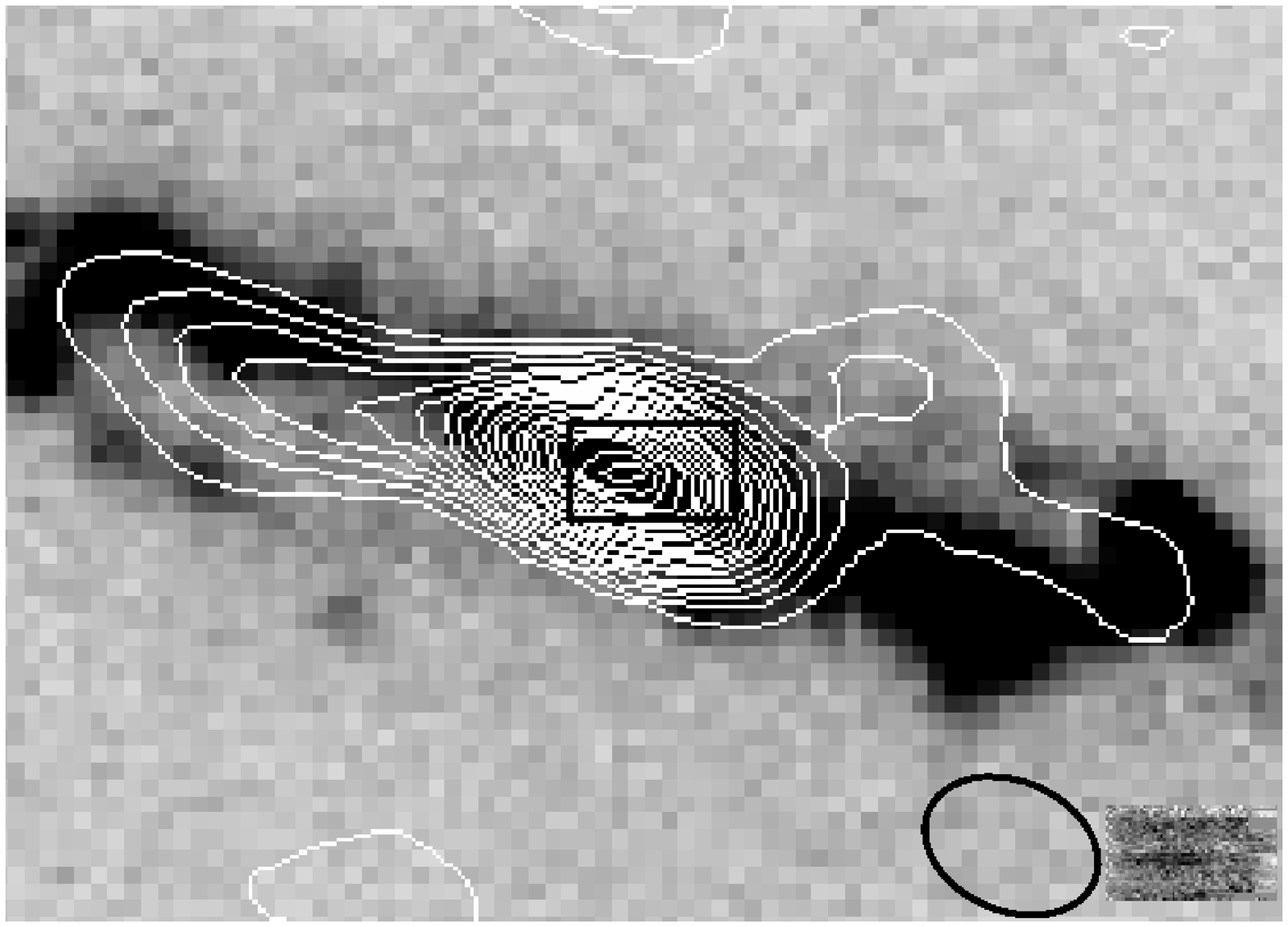}
\epsscale{1.0}
\caption[] {\ Similar to Figure~\ref{fig:CO_ha1} for NGC~6946 and NGC~7771.  The contours start at 15~Jy~beam$^{-1}$\kms\ for NGC~6946 and at 10~Jy~beam$^{-1}$\kms\ for NGC~7771, respectively, and are spaced by the same values.  The \COa\ and \hal\ maps for NGC~6946 are from Helfer et al. (2003) and Kennicutt et al. (2003), respectively.} 
\label{fig:CO_ha3}
\end{figure}

Contours of the \COa\ emission are overlaid on \hal\ maps in Figures~\ref{fig:CO_ha1}-\ref{fig:CO_ha3}.  Outlines of the regions mapped in \Htwo\ are included for reference.  The spatial profiles of \hal\ and CO emission for NGC~2388 and NGC~4418 are compact and centered on the nuclei.  In UGC~2855, NGC~6946, and NGC~7771 the \COa\ emission generally tracks the \hal\ distribution whereas NGC~1569 shows significant spatial differences in the two lines.  Additional details for the latter four extended sources are described below.  

\subsubsection{UGC 2855}
The \COa\ emission for UGC~2855 follows the spiral arm and bar patterns seen in \hal.  The molecular bar is $\sim$8~kpc long.  As described in H\"uttemeister, Aalto, \& Wall (1999), the radio continuum and \COa\ both peak within an arcsecond of 03$^{\rm h}$48$^{\rm m}$20.8$^{\rm s}$$+$70$\degr$07\arcmin58\arcsec.  The \hal\ and \Htwo\ emission also peak within an arcsecond of this position.  The northwestern part of the bar is a factor of 1.2-1.5 brighter in \COa\ than the southeastern portion (H\"uttemeister, Aalto, \& Wall).  The apparent linear cutoff along the northern edge is an artifact of the \COa\ map obtained by H\"uttemeister, Aalto, \& Wall.

\subsubsection{NGC 1569}
Taylor et al. (1999) have previously compared \COa\ emission with \hal\ for NGC~1569.  We have obtained \Htwo\ data for \HII\ region 2 of Waller (1991), which also overlaps with region C of Greve et al. (1996) and GMC 3 of Taylor et al. (1999).  Neither of the super starclusters for which NGC~1569 is famous is covered by the \Htwo\ map, though super starcluster A is near the eastern edge of the \Htwo\ map.  The \hal\ image exhibits fairly striking features due to the winds from the starclusters.  The \Htwo\ emission is strongest in the southwestern quadrant of the region mapped in \Htwo\ (Dale et al. 2004), a region that also shows significant \COa\ emission.

\subsubsection{NGC 6946}
Sheth et al. (2002) have previously compared the \COa\ emission with \hal\ for NGC~6946.  The \COa\ emission in the center of NGC~6946 displays a north-south structure and generally follows what is seen at \hal, though as pointed out by Sheth and collaborators, the \hal\ emission appears on the leading side of the \COa\ emission. 

\subsubsection{NGC 7771}
The linear structure to the \COa\ emission for NGC~7771 follows the general inclined-disk trend evidenced by \hal\ (and broadband optical and infrared images).  As for UGC~2855, the molecular disk is asymmetric with the northeastern side of the nucleus 50\% stronger in \COa\ emission than the southwestern.  Curiously, the molecular disk emission in the northeast is aligned with the \hal\ emission whereas the two distributions are largely misaligned for the southwestern portion of the disk.  The \hal\ morphology essentially follows that at 7\m\ (Dale et al. 2000), so the northeast-southwest differences cannot be explained by dust extinction.

\subsection{Flux}
The \COa\ fluxes integrated over the interferometer maps are listed in Table~\ref{tab:fluxes} along with fluxes extracted over the field-of-view of the Palomar Integral Field Spectrometer \Htwo\ pointings.\footnote{Note that the CO beams are $\sim$30\% larger than the PIFS fields-of-view for NGC~2388 and NGC~7771.}  In comparison with the global fluxes obtained from single-dish observations that are listed in the rightmost column of Table~\ref{tab:fluxes}, the integrated fluxes from the arrays are systematically lower.  Part of this discrepancy can be attributed to the relatively small \COa\ map sizes (for NGC~6946 and UGC~2855).  Some of the extended flux is also missed by the limited $uv$ coverage inherent to interferometric measurements; the minimum dish spacing of the OVRO ``compact'' configuration, for example, resolves out structures larger than $\sim35$\arcsec\ (see Helfer et al. 2003).  Detailed investigative work by Helfer et al. shows that, on 10\arcsec\ scales, interferometers capture 80-90\% of the total emission.  The recovery fraction approaches 100\% on these scales if the CO is dominated by a central concentration, similar to the situation for all our targets save for NGC~1569.

\subsection{Warm and Cold Molecular Hydrogen Masses}
Rigopoulou et al. (2002) find that the percentage of ``warm'' molecular gas as traced by mid-infrared \Htwo\ rotational lines ($T\sim150$~K), with respect to the bulk of the molecular gas that is ``cold'' and traced by the millimeter \COa\ transition, is 1-10\% by mass for starbursts, and 2-35\% for Seyferts.  The \Htwo\ mass at even higher temperatures ($T\sim2000$~K) can be computed from the extinction-corrected 2.122\m\ ro-vibrational \Htwo\ near-infrared line flux:
\be
M({\rm H}_2)_{\rm warm} = 2 m_{\rm p} n_{\rm H_2} V_{\rm H_2} = {2 m_{\rm p} F_{\nu=1\rightarrow0S(1)} 4\pi d^2 \over f_{v=1,J=3} A_{\nu=1\rightarrow0S(1)} h\nu } \simeq 5.08~{\rm M}_\odot \left(F_{\nu=1\rightarrow0S(1)} \over { 10^{-16}~\rm W~m}^{-2}\right) \left( d \over {\rm Mpc}\right)^2
\ee
where we assume the Orion vibrational temperature $T_{\rm vib}=2000$~K and thus a population fraction $f_{v=1,J=3}=0.0122$ (Scoville et al. 1982), and a 1$\rightarrow$0~S(1) transition rate of $A_{\nu=1\rightarrow0S(1)}=3.47\cdot$$10^{-7}$~s$^{-1}$ (Turner, Kirby-Docken, \& Dalgarno 1977).  
The ``cold'' molecular hydrogen mass can be inferred from the integrated \COa\ flux according to
\be 
M({\rm H}_2)_{\rm cold} \simeq 1.18 \cdot 10^{4}~{\rm M}_\odot \left(S_{\rm ^{12}CO(1\rightarrow0)} \over {\rm Jy~km~s^{-1}}\right) \left(d \over {\rm Mpc}\right)^2 \left(\alpha \over \alpha_{\rm G}\right) 
\ee
where $\alpha_{\rm G}$ is the Galactic \COa-to-\Htwo\ conversion factor (Wilson 1995).  The warm/cold \Htwo\ mass ratios range between 10$^{-7}$ to 10$^{-5}$ for the small fields mapped in \Htwo~2.122\m.

\section{DISCUSSION AND SUMMARY}
\label{sec:disc}
\begin{figure}[!ht]
\epsscale{1.0}
\plotone{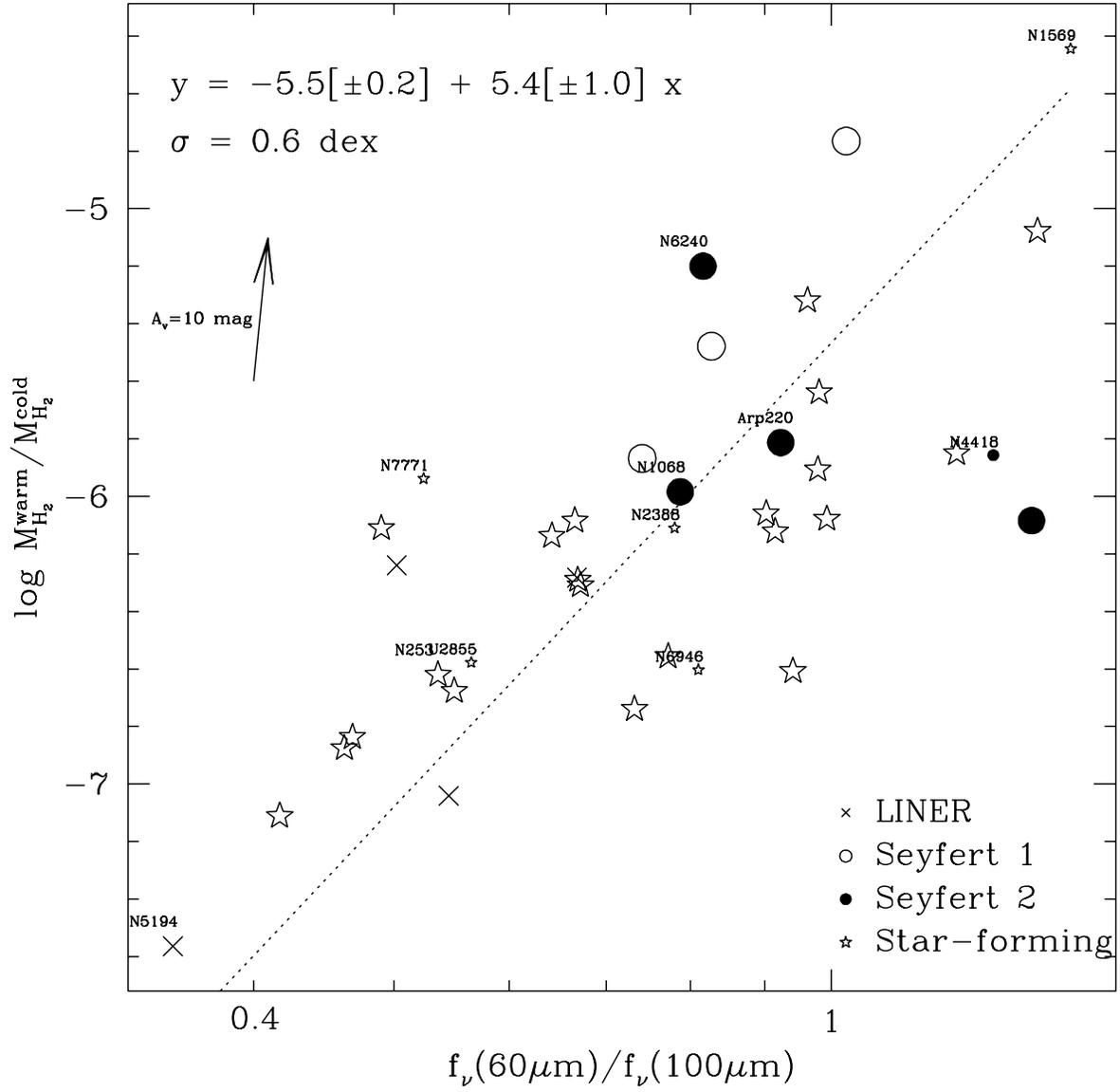}
\epsscale{1.0}
\caption[] {\ The ratio of warm-to-cold molecular gas mass as a function of far-infrared color.} 
\label{fig:ratios}
\end{figure}

The luminous infrared galaxy NGC~6240 has the largest known \Htwo~2.122\m\ line luminosity, and the largest ratio of \Htwo\ line-to-bolometric luminosity ratio ($L_{\rm H_2 2.122}\sim7.5\times10^7~L_\odot$, which is 0.3\% of the bolometric luminosity; see Draine \& Woods 1990).  The upper end of the range we find for our sample's {\it local} warm/cold molecular mass ratio (10$^{-7}$ to 10$^{-5}$) is consistent with the value for NGC~6240 (van der Werf et al. 1993; Young et al. 1995).  To expand upon our result and place it within a larger context, we have extracted a large pool of \Htwo~2.122\m\ and \COa\ data from the literature.  A large, uniform database of global \COa\ fluxes for nearby galaxies is provided by Young et al. (1995).  \Htwo~2.122\m\ data for at least 32 of these galaxies can be found in the literature, and for those galaxies with multiple measurements we use the fluxes from the largest apertures (Hall et al. 1981; Rieke et al. 1985; Heckman et al. 1986; Kawara et al. 1987; Fischer et al. 1987; Moorwood \& Oliva 1988; Puxley et al. 1988; Lester et al. 1990; Forbes et al. 1993; van der Werf et al. 1993; Calzetti 1997; Larkin et al. 1998; Roussel et al. 2003).  Since the \Htwo~2.122\m\ maps from the literature typically do not span areas as large as those covered by the Young et al. CO maps, we have estimated the portion of the CO flux that derives from within the \Htwo~2.122\m\ apertures.  We use an exponential model for the CO emission along with the scalelengths and effective CO radii suggested in Young et al.  The data are plotted in Figure~\ref{fig:ratios} as a function of global \IRAS\ far-infrared color.  The data are corrected for Galactic extinction and a uniform $\alpha/\alpha_{\rm G}=1$ is assumed.

The range in ``global'' warm/cold \Htwo\ mass ratio drawn from the literature spans a slightly larger range, extending down to 10$^{-7.5}$.  In addition, the mass ratio tracks the far-infrared color \IRAScolor.  Note also that the trend holds regardless of nuclear activity.  Correcting for internal extinction would likely steepen the slope of the trend, as sources with warmer far-infrared colors typically exhibit larger extinctions (e.g. Hattori et al. 2004).  However, compared to the $A_V\lesssim3$~mag typically found for starburst nuclei (Ho, Filippenko, \& Sargent 1997; Dale et al. 2004), the reddening vector displayed in Figure~\ref{fig:ratios} shows that a large internal extinction is necessary to significantly shift the data points.  Another caveat is that the CO-to-\Htwo\ conversion factor is likely to differ from galaxy to galaxy.  
Some authors suggest the CO-to-\Htwo\ conversion factor clearly depends on metallicity (e.g. Wilson 1995; Arimoto, Sofue, Tsujimoto et al. 1996; Israel 1997) whereas others claim negligible variation above a metallicity of 12+log[O/H]$\sim$8.0 (e.g. Rosolowsky et al. 2003; Bolatto et al. 2003).  Metallicities were found for half of the galaxies in the literature, and using the CO-to-\Htwo\ prescription described in (Wilson 1995) we find only a 10\% increase (0.04 dex) in the average warm/cold ratio.  
If internal extinction and the CO-to-\Htwo\ conversion factor $\alpha$ were properly gauged for each galaxy the data points would differ, but it is unlikely that the main trend would be washed out.
  
It is worth questioning whether this trend with \IRAScolor\ is just a global effect: Are the variations from the overall trend more significant on a local scale?  Does the observed trend simply reflect smoothing on a global scale of more variable local phenomena?   The \COa\ interferometric maps allow us to sample the cold molecular hydrogen on a scale more similar to the typical extent of the \Htwo~2.122\m\ emission.  The local far-infrared colors are estimated using \colore\ from the higher resolution \ISO\ mid-infrared imaging data and the \colore-to-\IRAScolor\ mapping for star-forming galaxies (Dale et al. 2001; Dale \& Helou 2002).  The small data points displayed in Figure~\ref{fig:ratios} represent the ``local'' ratios for the six galaxies in Table~\ref{tab:obs}---these data stem from just the Palomar Integral Field Spectrometer \Htwo\ fields-of-view.  Except for NGC~4418, which is listed as a Seyfert~2 galaxy in NED, the local ratios are indicated by stars to indicate that they derive from star-forming regions.  The data for these six small-scale (sub-kpc to kpc) regions are consistent with the galaxy global data.  The relation may break down on the parsec or tens of parsec scale (e.g. Scoville et al. (1982), where the mass ratio is 10$^{-3}-10^{-2}$ for the core of the Orion molecular cloud), but higher resolution data would be required to probe to this level in extragalactic regions.

The gas mass ratio gauges the molecular gas phase of the interstellar medium whereas the far-infrared color is a probe of the dust emission.  The two measures are coupled, likely due to the effects of star formation on the interstellar medium.  The warm/cold molecular gas mass ratio is a measure of star formation activity via shocks and perhaps UV excitation (Draine \& Woods 1990; van der Werf et al. 1993), while the far-infrared color is a measure of star formation activity via dust grain temperature (Hattori et al. 2004; Xilouris et al. 2004; Dale et al. in preparation).  In either case, the same stars are ultimately responsible for heating the warm molecular gas and interstellar dust.  A locally-based coupling could arise from molecule formation on dust or molecular gas excitation via the photoelectric effect from grains.  On the other hand, a more global description would simply link radiation leaking out from a younger stellar population embedded within or near a molecular cloud complex and heating interstellar dust distributed on kpc scales.  The photoelectric basis for local coupling is unlikely since photoelectrons primarily arise from the smallest grains, whereas the far-infrared color is dominated by emission from large grains (e.g. Hollenbach \& Tielens 1999).  Moreover, the far-infrared color is more of a large-scale measure since it reflects the balance of emission from diffuse cirrus and \HII/photodissociation regions (e.g. Helou 1986).  In other words, a global scaling is likely.  

We have obtained near-infrared \Htwo\ and interferometric \COa\ maps for six nearby galaxies, data that trace different phases of interstellar molecular gas.  We find that the mass ratio of warm and cold molecular gas is of order $10^{-7}$ to $10^{-5}$, indicative of the amount of molecular material that is experiencing the effects of active star formation.  After supplementing these sub-kpc to kpc scale data with integrated (global) data from the literature, we find a broad trend between the molecular gas mass ratio and the far-infrared color.  Both ratios are conceptually linked to star formation, though it is unlikely that they are directly physically coupled.  It is more likely that the link relates to radiation emerging from young stars within or near molecular clouds and heating dust grains distributed on kpc scales.

\acknowledgements 
We are thankful to C. Taylor for the use of his NGC~1569 CO map.  C. Kobulnicky, T. Jarrett, and E. Murphy provided useful comments and input.
The Owens Valley millimeter array is supported by NSF grant AST~99-81546.  This research has made use of the NASA/IPAC Extragalactic Database which is operated by JPL/Caltech, under contract with NASA.


\newpage
\scriptsize
\begin{deluxetable}{lcrccccc}
\tablenum{1}
\label{tab:obs}
\def\a{\tablenotemark{a}}
\def\b{\tablenotemark{b}}
\def\c{\tablenotemark{c}}

\def\x{$\times$}
\tablewidth{505pt}
\tablecaption{\COa\ Observations}
\tablehead{
\colhead{Galaxy} &
\colhead{R.A.~~~~~Decl.} &
\colhead{c$z$} & 
\colhead{$a_{\rm opt}$\x$b_{\rm opt}$}&
\colhead{Beam}               & 
\colhead{rms} & 
\colhead{Array}&
\colhead{Reference}
\\
\colhead{}       &
\colhead{(J2000)}        & 
\colhead{\tiny (km/s)}     & 
\colhead{}    &
\colhead{(\arcsec\x\arcsec)} &
\colhead{\tiny (mJy/beam)} & 
\colhead{}&
\colhead{}
}
\startdata
UGC~2855&034820.4~$+$700758&  1202&~4\farcm4\x2\farcm0&~5.4\x5.2&34 &OVRO&H\"uttemeister et al. (1999)\\
NGC~1569&043047.3~$+$645101&$-$104&~3\farcm6\x1\farcm8&~4.5\x3.9&7.5&IRAM&Taylor et al. (1999)\\
NGC~2388&072853.5~$+$334908&  4134&~1\farcm0\x0\farcm6&11.2\x7.3&14 &OVRO&This work\\
NGC~4418&122654.6~$-$005240&  2179&~1\farcm4\x0\farcm7&~5.5\x3.2&19 &OVRO&This work\\
NGC~6946&203452.3~$+$600914&  ~~48&11\farcm5\x9\farcm8&~6.0\x4.9&61 &BIMA&Helfer et al. (2003)\\
NGC~7771&235124.8~$+$200642&  4277&~2\farcm5\x1\farcm0&10.8\x7.8&13 &OVRO&This work\\
\enddata
\end{deluxetable}
\normalsize

\scriptsize
\begin{deluxetable}{lccccccc}
\tablenum{2}
\label{tab:fluxes}
\def\a{\tablenotemark{a}}
\def\b{\tablenotemark{b}}
\def\c{\tablenotemark{c}}
\def\d{\tablenotemark{d}}
\def\x{$\times$}

\tablewidth{480pt}
\tablecaption{Data}
\tablehead{
\colhead{Galaxy}&
\colhead{PIFS}  &
\colhead{CO}  &
\colhead{CO$_{\rm pifs}$\a}    &
\colhead{CO$_{\rm map}$\a}    &
\colhead{CO$_{\rm tot}$\b} &
\colhead{$M^{\rm warm}_{H_2}$/\c} &
\colhead{$f_\nu(60\micron)$/\d}
\\
\colhead{  }    &
\colhead{\Htwo\ map} &
\colhead{Map}&
\colhead{---}&
\colhead{(Jy km/s)}&
\colhead{---} &
\colhead{$M^{\rm cold}_{H_2}$} &
\colhead{$f_\nu(100\micron)$}
}
\startdata
UGC~2855&10\farcs0\x~5\farcs7&2\farcm6\x1\farcm1&190~~~~~~  &1200~~~~  &~2210  &$2.6\cdot10^{-7}$&0.57\\
NGC~1569&10\farcs0\x~5\farcs7&1\arcmin\x1\arcmin& ~0.86     &~~~~3.58  &~~~60  &$3.6\cdot10^{-5}$&1.5~\\
NGC~2388&10\farcs0\x~5\farcs7&2\arcmin\x2\arcmin& 38~~~     &116~~~    &\nodata&$7.8\cdot10^{-7}$&0.78\\
NGC~4418&10\farcs0\x~5\farcs7&2\arcmin\x2\arcmin& 50.~~     &~62~~~    &~~150  &$1.4\cdot10^{-6}$&1.3~\\
NGC~6946&10\farcs0\x10\farcs8&6\arcmin\x6\arcmin&1200~~~~~~~&9273~~~~~~&12370  &$2.5\cdot10^{-7}$&0.81\\
NGC~7771&10\farcs0\x~5\farcs7&2\arcmin\x2\arcmin&100~~~~    &337~~~    &~~540  &$1.2\cdot10^{-6}$&0.52\\
\enddata
\tablenotetext{a}{\footnotesize OVRO-based \COa\ fluxes are uncertain at the 20\% level.}
\tablenotetext{b}{\footnotesize FCRAO data (Young et al. 1995).}
\tablenotetext{c}{\footnotesize Mass ratios from the PIFS regions are uncertain at the 50\% level, driven by the large absolute uncertainty in the \Htwo~2.122\m\ fluxes (see Dale et al. 2004).}
\tablenotetext{d}{\footnotesize Infrared colors from the PIFS regions are uncertain at the 30\% level, based on a combination of uncertainties in the infrared fluxes and models (see \S~\ref{sec:disc} and Dale et al. 2000).}
\end{deluxetable}
\normalsize

\newpage
\end{document}